\documentclass[journal=jacsat,manuscript=article]{achemso}

\usepackage[version=3]{mhchem} 

\author{Maximilian Beddoe}
\email{maximilian.edmund.beddoe@uni-jena.de}
\affiliation[Friedrich Schiller University Jena]
{Institute of Solid State Physics, Friedrich Schiller University Jena, 07743 Jena, Germany}
\alsoaffiliation[Friedrich Schiller University Jena]
{Institute of Applied Physics, Abbe Center of Photonics, Friedrich Schiller University Jena, 07745 Jena, Germany.}
\author{Angela Barreda}
\affiliation[Friedrich Schiller University Jena]
{Institute of Solid State Physics, Friedrich Schiller University Jena, 07743 Jena, Germany}
\alsoaffiliation[Friedrich Schiller University Jena]
{Institute of Applied Physics, Abbe Center of Photonics, Friedrich Schiller University Jena, 07745 Jena, Germany.}
\alsoaffiliation[Carlos III University of Madrid]
{Group of Displays and Photonics Applications, Carlos III University of Madrid, Avda. de la Universidad, 30, Leganés, 28911 Madrid, Spain}

\author{Chengjun Zou}
\affiliation[Friedrich Schiller University Jena]
{Institute of Solid State Physics, Friedrich Schiller University Jena, 07743 Jena, Germany}
\alsoaffiliation[Friedrich Schiller University Jena]
{Institute of Applied Physics, Abbe Center of Photonics, Friedrich Schiller University Jena, 07745 Jena, Germany.}
\alsoaffiliation[Chinese Academy of Sciences]
{Institute of Microelectronics, Chinese Academy of Sciences, Beitucheng west road 3, Beijing, China}
\author{Daniel F\"ussel}
\affiliation[Friedrich Schiller University Jena]
{Institute of Applied Physics, Abbe Center of Photonics, Friedrich Schiller University Jena, 07745 Jena, Germany.}

\author{Muyi Yang}
\affiliation[Friedrich Schiller University Jena]
{Institute of Solid State Physics, Friedrich Schiller University Jena, 07743 Jena, Germany}
\alsoaffiliation[Friedrich Schiller University Jena]
{Institute of Applied Physics, Abbe Center of Photonics, Friedrich Schiller University Jena, 07745 Jena, Germany.}

\author{Dmitry Pidgayko}
\affiliation[Friedrich Schiller University Jena]
{Institute of Solid State Physics, Friedrich Schiller University Jena, 07743 Jena, Germany}
\alsoaffiliation[Friedrich Schiller University Jena]
{Institute of Applied Physics, Abbe Center of Photonics, Friedrich Schiller University Jena, 07745 Jena, Germany.}

\author{Thomas Pertsch}
\affiliation[Friedrich Schiller University Jena]
{Institute of Applied Physics, Abbe Center of Photonics, Friedrich Schiller University Jena, 07745 Jena, Germany.}
\alsoaffiliation[IOF]
{Fraunhofer Institute for Applied Optics and Precision Engineering, Albert-Einstein-Str. 7, 07745 Jena, Germany}
\author{Isabelle Staude}
\email{isabelle.staude@uni-jena.de}
\affiliation[Friedrich Schiller University Jena]
{Institute of Solid State Physics, Friedrich Schiller University Jena, 07743 Jena, Germany}
\alsoaffiliation[Friedrich Schiller University Jena]
{Institute of Applied Physics, Abbe Center of Photonics, Friedrich Schiller University Jena, 07745 Jena, Germany.}
\title[An \textsf{achemso} demo]
   {Dynamic tuning of anisotropy-induced quasi-bound-states-in-the-continuum in liquid-crystal integrated dielectric metasurfaces}

\abbreviations{IR,NMR,UV}
\keywords{American Chemical Society, \LaTeX}

\begin{document}

\begin{abstract}
Symmetry-protected quasi-bound states in the continuum (qBICs) in all-dielectric metasurfaces can be excited either by introducing a geometrical asymmetry into the unit cell or by illuminating the metasurface under oblique incidence. To date, tuning the coupling strength of such qBICs to the continuum has mainly relied on control of the geometrical asymmetry. Here, we show that qBICs can also be induced in a symmetric dielectric metasurface by embedding the latter into an anisotropic birefringent medium, and that they can be actively tuned by either rotating the anisotropy axis of the medium or by varying the degree of anisotropy. We experimentally implement a corresponding system by embedding a highly symmetric silicon nanocylinder metasurface into a nematic liquid crystal (LC) cell.
Importantly, the LC does not only provide the desired birefringence, but also dynamic control of its strength as well as of the orientation of its anisotropy axis in response to different external stimuli. Specifically, applying voltage across the LC cell, where the LC molecules are prealigned along one of the lattice directions of the metasurface, enables out-of-plane rotation of the anisotropy axis of the LC medium. Meanwhile, its birefringence can be tuned by changing the temperature. In our experiments, the coupling of the qBICs to the radiation continuum is first maximized by the application of 25 V across the LC cell at room temperature. We then systematically tune the temperature and show that the resonance depth of the qBIC modes, which is associated with their coupling strength to the continuum, of the qBIC modes increases monotonously with birefringence. 
Our findings establish anisotropy-induced qBICs in LC integrated dielectric metasurfaces as a promising platform for actively reconfigurable metasurfaces and their applications in displays, optical modulation and tunable photonic devices.
  
\end{abstract}
\section{Keywords}
Tunable metasurfaces, active tuning, liquid crystals, Mie-resonances, all-dielectric nanophotonics, quasi-bound states in the continuum
\section{Introduction}
All-dielectric metasurfaces are two-dimensional arrangements of subwavelength dielectric scatterers that enable unprecedented control over light beyond the capabilities of conventional optical components \cite{chen2016review, LIU2020101868}.
Their functionality arises from the ability to engineer the optical response of individual scatterers, often called meta-atoms, by tailoring their geometry \cite{Kuznetsov2016, Decker2015}, arrangement \cite{rahimzadegan2019disorder, gutruf2016mechanically}, constituent material \cite{Kuznetsov2016} and the refractive index of the surrounding environment \cite{LiuXu2020, XuJ2021}.
 On the basis of their tailorable optical responses, optical metasurfaces have emerged as a versatile platform for realizing compact and highly functional flat optical devices. \cite{Yu2014, Kuznetsov2024, Schulz2024}
Various applications including lidars \cite{yang2022active, juliano2022metasurface}, metalenses \cite{yang2022active, Aieta2012, Lin2014, Decker2019}, augmented reality \cite{yang2022active, song2025full, aththanayake2025tunable}, lasing \cite{bashiri2026spatio, do2025room} and beam steering \cite{yang2022active, qiu2025wireless} were demonstrated.
Many of these applications require active tunability or reconfigurability, which has been achieved using, e.g., phase change materials \cite{santonocito2023recent, barreda2022}, photo-generated charge carriers in semiconductors \cite{santonocito2023recent, shcherbakov2017ultrafast}, responsive polymers \cite{santonocito2023recent, Walden2024, Zou2023, Barkey2024}, and liquid crystals (LCs) \cite{Sautter2015, Komar2017, Zou2019, Zou2021, Beddoe2025}.
LCs are highly promising materials in this context due to their application-relevant switching times \cite{jiao2008alignment} and operating parameters, widespread industrial deployment, and multiresponsivity \cite{Jones2002, bedowr2024crystalline}.  

At room temperature, most LCs are in the nematic phase, in which its elongated molecules exhibit a long-range orientational order while lacking a long-range positional order \cite{Jones2002}. 
Consequently, nematic LCs are optically anisotropic and can be described by an uniaxial refractive index tensor, whose optic axis is defined by the average molecular orientation (the director).

Importantly, the refractive index of the LCs varies as a function of temperature \cite{Li2005}. Moreover, when the temperature increases beyond a material-specific critical temperature T$_c$, they undergo a phase transition to the isotropic phase. In this state, the long-range orientational order is lost, and molecules adopt random orientations \cite{Jones2002}, causing the LC refractive index to become isotropic. Based on these effects, temperature can be used as an effective control parameter for LC integrated metasurfaces \cite{Sautter2015, Zou2021}. 
Another important tuning mechanism involves the reorientation of the anisotropy axis of the LC via an external voltage \cite{Jones2002, alex2019analytical, Komar2017}, light \cite{Beddoe2025, yang2026all, izdebskaya2026all} and magnetic fields \cite{Izdebskaya2022, Izdebskaya2023}, which consequently modulates the optical response of the LC-integrated metasurface.
The integration of metasurfaces into LCs has enabled a variety of tunable photonic devices, including dynamically adjustable color generators and gradients \cite{Badloe2022}, spatial light modulators \cite{Moitra2023}, advanced display technologies \cite{LiLiu2020, Ou2025LCoS}, and tunable metalenses \cite{Bosch2025metalense}.
However, most works so far exploited the mere change in the effective refractive index of the LC material surrounding the meta-atoms for the optical modes of interest, thereby neglecting the unique potential offered by the powerful concept of symmetry -- and tailored symmetry breaks -- in LC integrated metasurfaces. 
Such symmetry considerations are particularly relevant in the context of symmetry-protected quasi-bound states in the continuum (qBICs), where the optical response is directly governed by breaking the structural symmetry of the metasurface system \cite{barreda2025bound, koshelev1, koshelev2018}. 
Conventionally, symmetry breaking in qBIC metasurfaces is induced by altering the geometry of the metasurface through perturbations in the shape and/or arrangement of the meta-atoms. \cite{PhysRevA.100.063803, barreda2025bound, LIU2024111106}.
Alternatively, qBICs can be excited in C$_{4v}$-symmetric disk-based metasurfaces by tilting either the entire metasurface \cite{murai2020bound, hidalgo2025cloaked} or the individual nanoresonators \cite{hidalgo2025cloaked}. 
Recently, an approach based on permittivity-induced symmetry breaking has also been emerging in which the refractive index of the meta-atoms or their surroundings is locally varied. \cite{barreda2025bound, yang2025permittivity, Berte2023}.
The coupling strength of the qBIC to the radiation directly correlates with the symmetry-breaking \cite{koshelev2018}. A quantitative way to determine the coupling strength is to calculate the radiative Q-factors of the resonances. \cite{koshelev2018}
Considering that the refractive index of the composite material of a nanostructure can typically be changed in a controlled and reversible fashion much easier than its geometry, a significant advantage of this approach is the possibility of actively controlling the coupling strength of the qBIC resonances to the continuum using various external stimuli.
For example, optical pumping has been previously used to locally vary the refractive index of the meta-atoms via photoexcitation of carriers, leading to the demonstration of ultrafast tuning of the qBIC modes \cite{Aigner2025, yang2024ultrafast}. However, major limitations of this approach are the relatively short lifetime of the photoexcited carriers \cite{peter2010fundamentals} and small accessible refractive index changes. Moreover, the temporal evolution of the resonance is dictated by the intrinsic carrier dynamics after photoexcitation, limiting the degree of active control. Furthermore, the coupling strength remained constrained by the overall asymmetry of the investigated metasurfaces in the demonstrations until now.
As another example of permittivity-induced symmetry breaking, electrical reconfigurability of the asymmetry of the meta-atom environment has been implemented by locally applying an isotropic conductive polymer as a surrounding material \cite{hu2024environmental}. The primary drawback of this approach lies in the significant fabrication complexity of such structures.
Overall, in most qBIC implementations, the coupling strength has remained intrinsically linked to the metasurface asymmetry.
Decoupling the coupling strength from the structural asymmetry would simplify the fabrication process and eliminate the need for complex tuning strategies.
Remarkably, the opportunities for introducing tailored symmetry breaks to symmetric metasurfaces, which are provided by intrinsically anisotropic birefringent materials in general and nematic LCs in particular, have been largely overlooked so far. 
While LCs have previously been used for tuning of metasurfaces supporting qBICs, prior work mainly focused on resonance wavelength changes \cite{Izdebskaya2022, yu2025liquid, Manian2025, Chang2025, zhang2025momentum}.
Also, it was numerically shown that symmetry-protected qBICs can be suppressed by LCs, when the latter has the same effective refractive index as the material of the metasurface \cite{zhang2026liquid}. 
In contrast, the concept of material-anisotropy-induced qBICs was not systematically exploited in metasurfaces so far. Specifically, the generation and tuning of qBIC modes in dielectric metasurfaces based on dynamically induced symmetry breaks associated with the anisotropic response of nematic LC remains unexplored.
Anisotropy-induced photonic qBICs have previously been discussed for waveguide structures containing anisotropic birefringent materials, which have been shown to support qBICs with a range of interesting and exotic properties \cite{gomis2017anisotropy}. Tunable waveguide structures incorporating LCs have also been demonstrated in this context \cite{baral2025dynamic}.
Furthermore, it was numerically demonstrated that LCs can collectively excite and tune different types of surface lattice resonances. \cite{habibpourmoghadam2026anisotropy}.
Here, we experimentally demonstrate the dynamic generation and tuning of anisotropy-induced photonic qBICs in symmetric dielectric metasurfaces integrated in a nematic LC cell, where the LC acts as a multiresponsive anisotropic birefringent material. Importantly, the multiresponsivity of the LC \cite{Zou2021} allows both for actively inducing discrete symmetry breaks starting from a highly symmetric configuration, enabling to “switch on” a qBIC state, and for continuous asymmetry factor tuning. The latter provides dynamic control over the coupling strength to the continuum for a given qBIC mode, resulting, e.g., in tuning of its Q-factor or resonance depth. Our experimental results are supported by full-wave numerical calculations, including a multipolar analysis of the excited qBIC resonances.  
Our work highlights the potential of tailored symmetry breaks provided by birefringent materials for the study of anisotropy-induced photonic qBICs in metasurfaces and establishes nematic LCs as a viable material choice for their dynamic control.

\section{Results and Discussion}

\begin{figure}
\centering
\includegraphics[trim=1 1 1 1,clip,width=\textwidth]{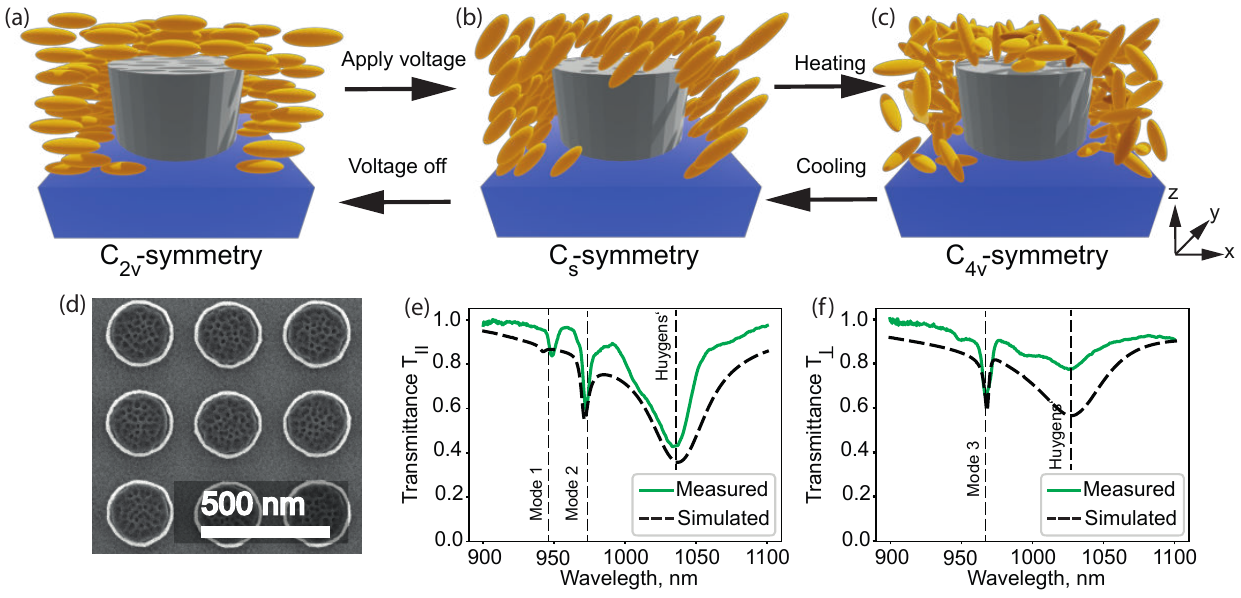}
  \caption{Schematic illustration of the three different cases considered regarding the state of the LC surrounding the meta-atoms and thereby the overall symmetry of the hybrid LC-metasurface unit cell: (a) C$_{2v}$-symmetry, (b) C$_{s}$-symmetry and (c) C$_{4v}$-symmetry. (d) Top-view SEM image of the metasurface. Measured and simulated transmittance spectra of the fabricated metasurface embedded in a LC cell for incident light polarized (e) parallel ($T_{\parallel}$) and (f) perpendicular ($T_{\perp}$) to the LC orientation.}
  \label{fig1}
\end{figure}

A conceptual sketch of our experiment is shown in Figures \ref{fig1} (a-c). A symmetric dielectric metasurface composed of silicon nanocylinders arranged on a square lattice is embedded into a nematic LC cell. 
The sample was fabricated by first sputtering an indium tin oxide (ITO) layer, then depositing an SiO$_2$-layer with thermal evaporation, and finally a silicon layer by plasma-enhanced chemical vapour deposition on a SiO$_2$-substrate. The ITO is used as a transparent electrode to apply voltage across the LC cell. The electrical contact areas of the ITO layers were created using UV lithography to mask the ITO layer during subsequent deposition steps. Finally, the metasurface pattern was defined using electron beam lithography (EBL) in combination with reactive ion etching of the silicon layer (see Ref. \cite{Zou2021} for details on the fabrication procedure). 
The resulting nanocyclinders have a radius of $r=179$\,nm and a height of $h=166$\,nm. The period is $a=520$\,nm. The silicon nanocylinders are each situated on an SiO$_{2}$ pedestal with a height of $h_{e}=53$\,nm resulting from slight over-etching.
For integration of the fabricated metasurface into a nematic LC cell, we follow the procedure described in the Supporting Information, section S6. Three different cases are then considered regarding the state of the LC surrounding the meta-atoms, and, consequently, the overall symmetry of the hybrid LC-metasurface system. These are illustrated in Figure \ref{fig1} (a-c), respectively.  We hypothesize that the possibility to observe qBIC modes should depend on the symmetry of the system, where lowering the symmetry is expected to increase the opportunities for specific symmetry-forbidden modes to couple to the continuum.
To prealign the LCs homogeneously in $x$-direction \cite{Beddoe2025}, the metasurface and an SiO$_{2}$ coverplate were spin-coated with a thin layer of the photoalignment material AmA2522 \cite{Muravskyama2522}  with a thickness $h_{pa}$ of 10~nm . After assembling the LC cell (details see in Supporting Information section S7) and injecting the nematic LC (Licristal E7 from Merck), the AmA2522 layers were exposed to $y$-polarized 475\,nm cw laser light (Source: Photontec 457nm DPSS) ($12.3$~W/cm${^{2}}$) for 20 minutes. During the exposure, the LC was heated using a heating resistor to the isotropic state (see Ref. \cite{Beddoe2025} for details on the setups and the procedure of the prealignment of the LCs and Ref. \cite{Zou2019} for the details of the LC cell). 
The cell thickness was measured to be 8 $\mu$m using linear-optical spectroscopy and fitting of the observed Fabry-Perot fringes before LC infiltration.
As a result of the described procedure, the meta-atoms are embedded into an anisotropic birefringent material with its anisotropy axis oriented along $x$-direction. In this first case considered illustrated in Fig. 1 (a), the overall point group symmetry of the metasurface unit cell corresponds to a C$_{2v}$-symmetry \cite{haken2013molecular}. Note, however, that imperfections in the prealignment, which are commonly observed in experiments, can be expected to introduce deviations from perfectly symmetric conditions \cite{Dolan2021}.
Figure \ref{fig1} (b) displays the second case, where the LC remains in its nematic state. Here, we assume the application of a voltage across the LC cell, leading to an out-of-plane rotation of the LC molecules. The rotation angle, i.e. the angle between the LC director and the $x$-axis (see inset in Figure \ref{fig1} (b)), depends on the magnitude of the applied voltage and on the z-position within the LC cell. This rotation reduces the overall point group symmetry of the hybrid LC-metasurface system to a C$_s$-symmetry \cite{haken2013molecular}, since the mirror symmetry with respect to the $y-z$ plane is now broken.\\
Finally, Figure \ref{fig1} (c) illustrates the third case considered, where the LC is heated above its critical temperature and consequently transitions to its isotropic phase. This case represents the most symmetric of the system states considered, corresponding to the C$_{4v}$ -symmetry \cite{haken2013molecular}. Moreover, as here the symmetry condition does not rely on perfect prealignment, this case can be expected to be experimentally more robust than the first case.
Before progressing with a systematic analysis of the three considered cases, we confirm the successful fabrication and LC integration of our metasurface for the experimentally least complex first case of room temperature without any external stimulus applied.
A scanning electron microscopy (SEM) image of the fabricated sample is shown in Fig. \ref{fig1} (d).
Experimentally measured and numerically simulated spectra of the metasurface before LC integration are included in section S1 of the Supporting Information. 
The transmittance of the LC-integrated metasurface was measured using the optical setup shown in Figure S5 of the Supporting Information section S5.
Corresponding results for the incident light polarized parallel ($T_{\parallel}$, x-polarized) and perpendicular ($T_{\perp}$, y-polarized) to the LC-prealignment direction are shown in Fig. \ref{fig1} (e,f), respectively.
For $T_{\parallel}$, two qBICs, termed Modes 1 and 2 in the following, and for $T_{\perp}$ one qBIC, termed Mode 3 in the following, can be observed. On the long-wavelength side of Modes 1 to 3, a quasi Huygens' regime of partially overlapping electric and magnetic dipolar Mie-type resonances \cite{Decker2015} is observed at around 1040~nm wavelength for both polarizations as in accord with the literature \cite{Zou2021}.
To compare our measured results with theory, we performed numerical calculations using the frequency-domain solver of the commercial software package CST Studio Suite (see Section S2 in SI for details). For silicon, dispersive refractive index data obtained from ellipsometry measurements on the unstructured films is used.
The refractive index of the glass substrate is set to 1.45 \cite{malitson1965interspecimen}. Losses and the buried ITO layer are neglected. 
The LC is modelled as an anisotropic material with the ordinary refractive index $n_o$ and extraordinary refractive index $n_e$. Their dispersion is calculated using the formalism from Li et al. \cite{Li2005}
Note that the agreement with the experimental spectra in LC-infiltrated metasurfaces is commonly improved, if the impact of the complex nanoscale geometry on prealignment quality is considered \cite{Veetil2024, jiao2008alignment}. While precise modeling of the LC director is cumbersome \cite{decker2013electro}, adjusting the rotation angle in an ad-hoc fashion in the region between the meta-atom was previously shown to allow for effective modelling of the LC-integrated metasurface response. 
In our case, to optimize the agreement of the numerically calculated spectra with experimental data, we considered a rotation angle of $\Theta_{P1} = 38 ^\circ$, the so-called pretilt angle, in the region between the meta-atoms to effectively account for alignment complexities in the nanoscale topography to optimize the agreement between the measurements and the simulations \cite{Zou2019, zhang2025momentum} (see refined model shown in Fig. S2(b) and text in Section S2 in Supporting Information for explanation). Importantly, this adjustment entails a symmetry break in the hybrid LC-metasurface system.
The calculated spectra are displayed along the corresponding experimental results in Fig. \ref{fig1} (e,f).
A good qualitative agreement is observed between the measured and calculated transmittance spectra. The remaining deviations between the measured and simulated transmittance spectra can be explained by fabrication \cite{Kuehne2021} and LC orientation imperfections \cite{Dolan2021}. 

\begin{figure}
\centering
\includegraphics[trim=1 1 1 1,clip,width=\textwidth]{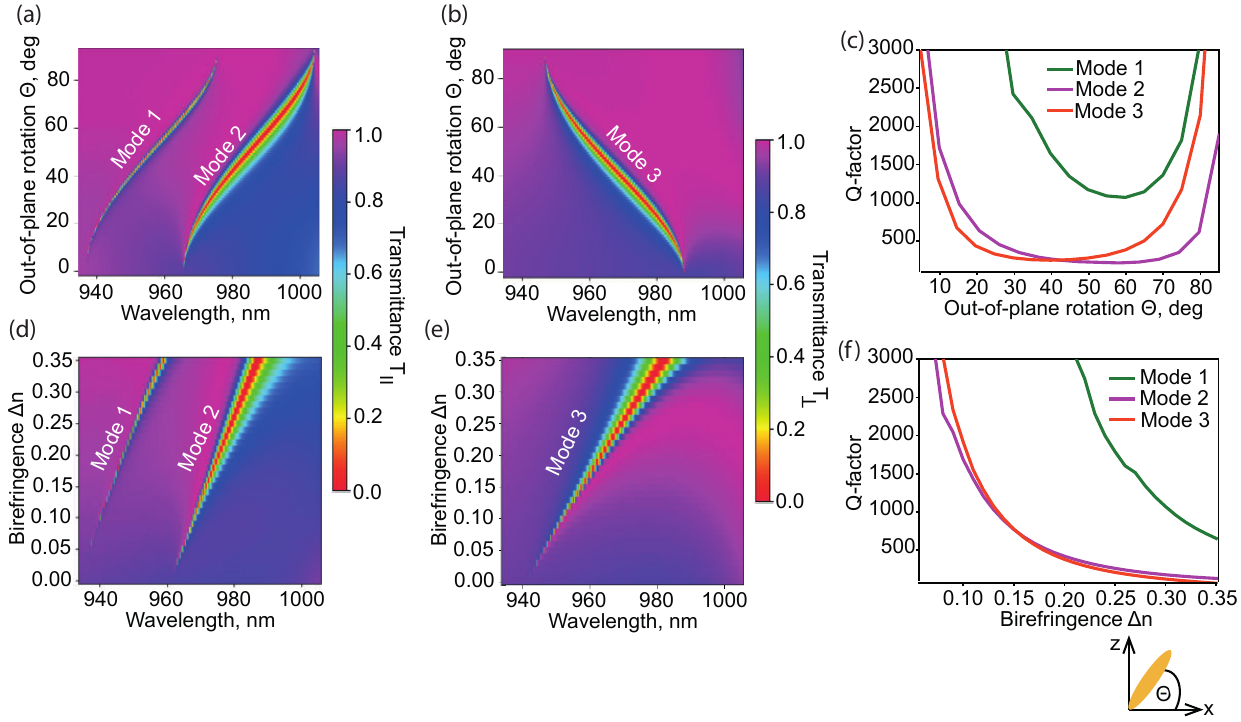}
  \caption{Numerically calculated transmittance spectra (a) $T_{\parallel}$ and (b) $T_{\perp}$ for out-of-plane rotation $\Theta$ ranging from 0 degrees to 90 degrees for LC with ordinary  $n_o=1.5$ and extraordinary refractive index $n_e=1.75$. (c) Q-factors of Modes 1 to 3 as a function of out-of-plane rotation $\Theta$. Simulated transmittance spectra (d) $T_{\parallel}$ and (e) $T_{\perp}$ for with a systematic variation of the LC birefringences $\Delta$n from $\Delta n=0.0$ to $\Delta n=0.3$. Here, the ordinary refractive index is fixed to $n_o=1.5$ and the out-of-plane rotation to $\Theta=40^\circ$. (f) Q-factors of modes 1 to 3 depending on the birefringence $\Delta n$}
  \label{fig2}
\end{figure}

Next, we proceed with a systematic investigation of the influence of the three different symmetry conditions on the optical response of the LC integrated metasurface, paying particular attention to the possibility to excite qBIC states. Initially, to this end, we perform numerical calculations on a more conceptual level, neglecting experimental complexities. Specifically, in contrast to the above-described results, we neglect the pretilt angle in the region between the meta-atoms, the local variation of the LC orientation and the fabrication imperfections (for further explanation of model, see ideal model shown in Figure S2(a) and text in Section S2 in the Supporting Information).
Under these conditions, we first concentrate on cases one and two (Figure \ref{fig1} (a,b)) by systematically investigating the effect of the global out-of-plane rotation $\Theta$ of the LC anisotropy axis on the optical response starting from the highly symmetric case one. Corresponding results are shown in Figure \ref{fig2}.

In accord with Fig. \ref{fig1} (e,f), at finite out-of-plane rotations $\Theta$ for $T_{\parallel}$ (Fig. \ref{fig2} (a)), two modes, Modes 1 and 2, and for $T_{\perp}$ (Fig. \ref{fig2} (b)) one mode, Mode 3, can be observed. All modes show a similar behaviour. 
Importantly, they all vanish at out-of-plane rotations of $\Theta=0^{\circ}$ and $\Theta=90^{\circ}$ corresponding to the configurations of highest symmetry.  They also become broader around  $\Theta=45^{\circ}$
To quantify this broadening effect, their Q-factors are displayed as a function of out-of-plane rotation $\Theta$ in Figure \ref{fig2} (c). 
The minima of the Q-factors of Modes 1 to 3 as a function of $\Theta$ are observed at $\Theta=60^{\circ}$, $55^{\circ}$ and $40^{\circ}$, respectively. $Q$ diverges for Modes 1 to 3 when approaching $\Theta \to 0^\circ$ and $\Theta \to 90^\circ$.
These observations are consistent with the hypothesis that the out-of-plane rotation of the LC molecules and the associated rotation of its anisotropic refractive index tensor to an oblique angle induces a symmetry-break in the metasurface, which allows the qBIC modes to couple to the radiation continuum.
 At $\Theta = 0^\circ$ and $\Theta = 90^\circ$, the hybrid LC-metasurface system exhibits high symmetry (C$_{2v}$-symmetry at 0$^\circ$ and C$_{4v}$-symmetry at 90$^\circ$ \cite{haken2013molecular}), whereas the symmetry is reduced to a C$_{s}$-symmetry point group for oblique angles.
Next, we concentrate on the highly asymmetric case of $\Theta=40^\circ$ and investigate, how changing the birefringence $\Delta n=n_e-n_o$ affects Modes 1 to 3. Including the case of vanishing $\Delta n=0$ in our analysis allows us to expand the discussion also to the highly symmetric case three (Fig. \ref{fig1} (c)).
Figures \ref{fig2} (d-e) shows the corresponding transmittance spectra of the LC-integrated metasurface for the incident light polarized parallel and perpendicular to the LC prealignment direction, respectively, and for a systematic variation of the birefringence $\Delta n$ ranging from 0.0 to 0.3 with $n_o=1.5$. For $\Delta n=0$, all qBIC modes disappear as expected from the above discussion. For finite $\Delta n$, the linewidth of all qBIC modes increases systematically with $\Delta n$ associated with the corresponding decrease in Q-factor as shown in Figure 2 (f). 
These results indicate that the symmetry breaks introduced by the anisotropy of the LC is responsible for allowing Modes 1-3 to couple to the radiation continuum, where the strength of the symmetry breaking determines the coupling strength.

\begin{figure}
\includegraphics[trim=1 1 1 1,clip,width=\textwidth]{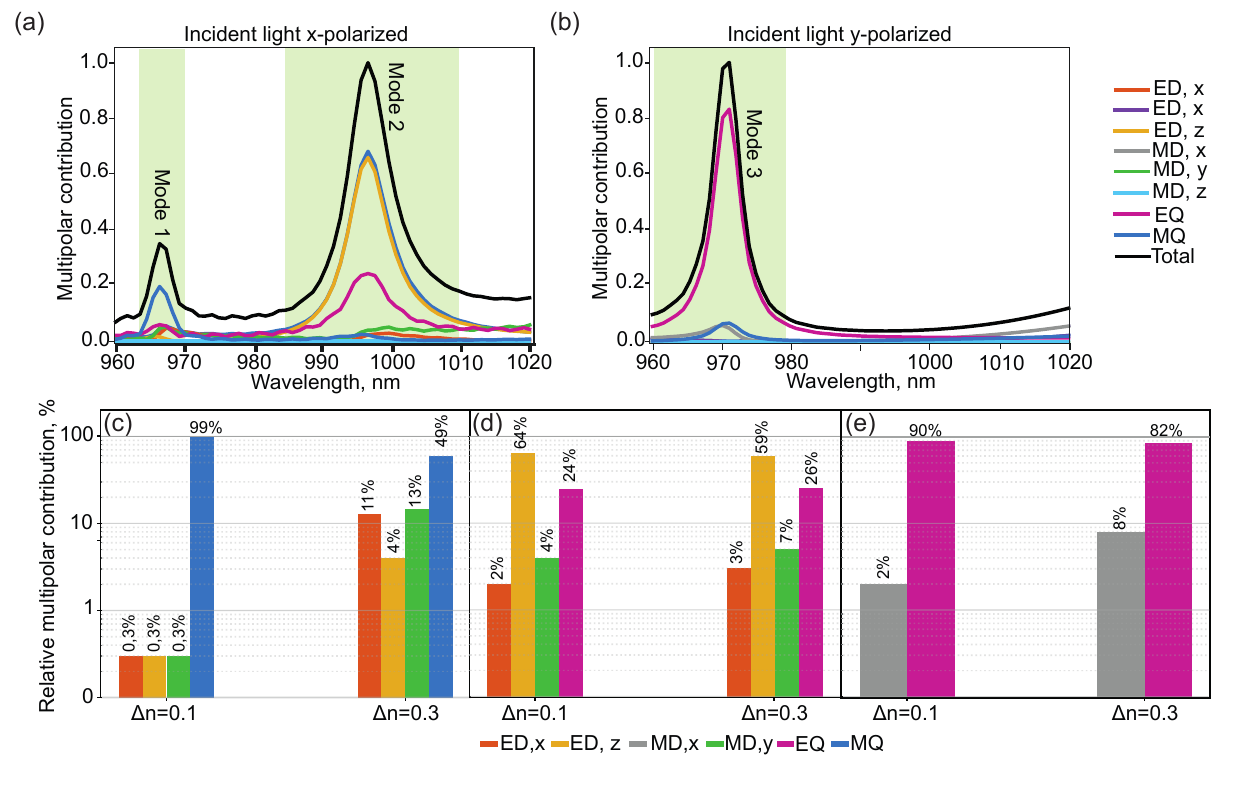}
\caption{Multipolar contributions to the total radiated power for an out-of-plane rotation of the LC anisotropy axis of $\Theta=40^{\circ}$:  Incident light polarized (a) parallel and (b) perpendicular to LC prealignment direction for a birefringence of $\Delta n=0.3$. The radiated power of each multipole in (a,b) is normalized to the spectral maximum of the total radiated power. Relative contributions of the different multipoles to the total power radiated by (c) Mode 1, (d) Mode 2 and (e) Mode 3 for birefringence values of $\Delta n = 0.1$ and $\Delta n = 0.3$. The green rectangles in (a,b) indicate the spectral regions corresponding to Modes 1–3.}
\label{fig3}
\end{figure}

In order to analyze the emergence of the qBIC modes to the radiation continuum for the reduced-symmetry cases in more detail \cite{PhysRevA.100.063803}, the multipolar decomposition was performed using the finite-difference time-domain (FDTD) method implemented in Ansys Lumerical \cite{he2018toroidal}.
Note that although the local current distributions are calculated in the periodic metasurface arrangement, we analyze the radiation power arising from the current distributions excited within individual meta-atoms radiating into free space. The radiated power associated with each multipolar contribution was calculated from the induced current distribution within the unit cell. 
The resulting multipolar spectra were then normalized to the maximum value of the spectrum obtained by summing the radiated powers of all multipoles considered in the decomposition to obtain the multipolar contributions to the total radiated power.
Figure \ref{fig3} (a, b) shows the multipolar contributions to the total radiated power for incident light polarized parallel (x-pol) and perpendicular (y-pol) to the LC orientation, respectively, again considering the highly asymmetric case of $\Theta=40^\circ$. 
For Mode 1, the dominant contribution is a magnetic quadrupole (MQ), which is nonradiative due to a symmetry mismatch with the incident plane wave \cite{Sadrieva2019multipolar}. The coupling of the qBIC to the radiation continuum originates from the electric dipole in x-direction (ED,x) and magnetic dipole in y-direction (MD,y).
Mode 2 is dominated by an electric dipole mode in z-direction (ED,z) and a smaller but significant electric quadrupole (EQ) contribution, which are both nonradiative in normal direction.
Coupling to the far-field of Mode 2 becomes possible through small contributions from electric dipoles in x-direction (ED,x) and magnetic dipoles in y-direction (MD,y).
Mode 3 has an electric quadrupole (EQ), a magnetic quadrupole (MQ) and a magnetic dipole (MD,x) contribution.
To investigate the evolution of the far-field response for Modes 1 to 3 with increasing anisotropy-mediated asymmetry, the relative multipolar contributions to the total radiated power are compared for birefringence values of $\Delta n=0.1$ and $ \Delta n=0.3$ (Fig. \ref{fig3} (c-e)). 
The relative multipolar contributions to the total power radiated by Modes 1-3 are calculated by dividing the integral of the radiated power of a single multipole over the spectral range of the corresponding mode through the integral of the sum of the radiated powers of all multipoles (black curves in Fig. \ref{fig3} (a,b)) over the same spectral range. 
The spectral ranges for Modes 1 to 3 are indicated with the green rectangles in Fig. \ref{fig3} (a,b).
Note that $\Delta n = 0$ cannot be included in the analysis since no qBICs are excited for this case. In Mode 1, the dipolar contributions ED,x and ED,z increase with birefringence, whereas the MQ contribution decreases. For Mode 2, the contributions from ED,z and EQ do not present any notable change, whereas the contribution from ED,x increases. In Mode 3, the small contributions of (MD,x) and (MD,z) are enhanced, while the EQ contribution is reduced.
The multipolar analysis is fully consistent with the observed features from the corresponding calculated near-field profiles of Modes 1 to 3 (see Supporting Information, Figure S3).
Altogether, recalling that out-of-plane far-field coupling relies on nonvanishing in-plane-oriented ED and/or MD contributions (i.e., ED,x and MD,y for x-polarized far-fields and ED,y and MD,x for y-polarized far-fields) \cite{murai2020bound}, our analysis clearly reveals the mechanisms behind the observed decrease of the Q-factor with growing anisotropy presented in Fig. \ref{fig2} (e).

\begin{figure}
\centering
\includegraphics[trim=1 1 1 1,clip,width=\textwidth]{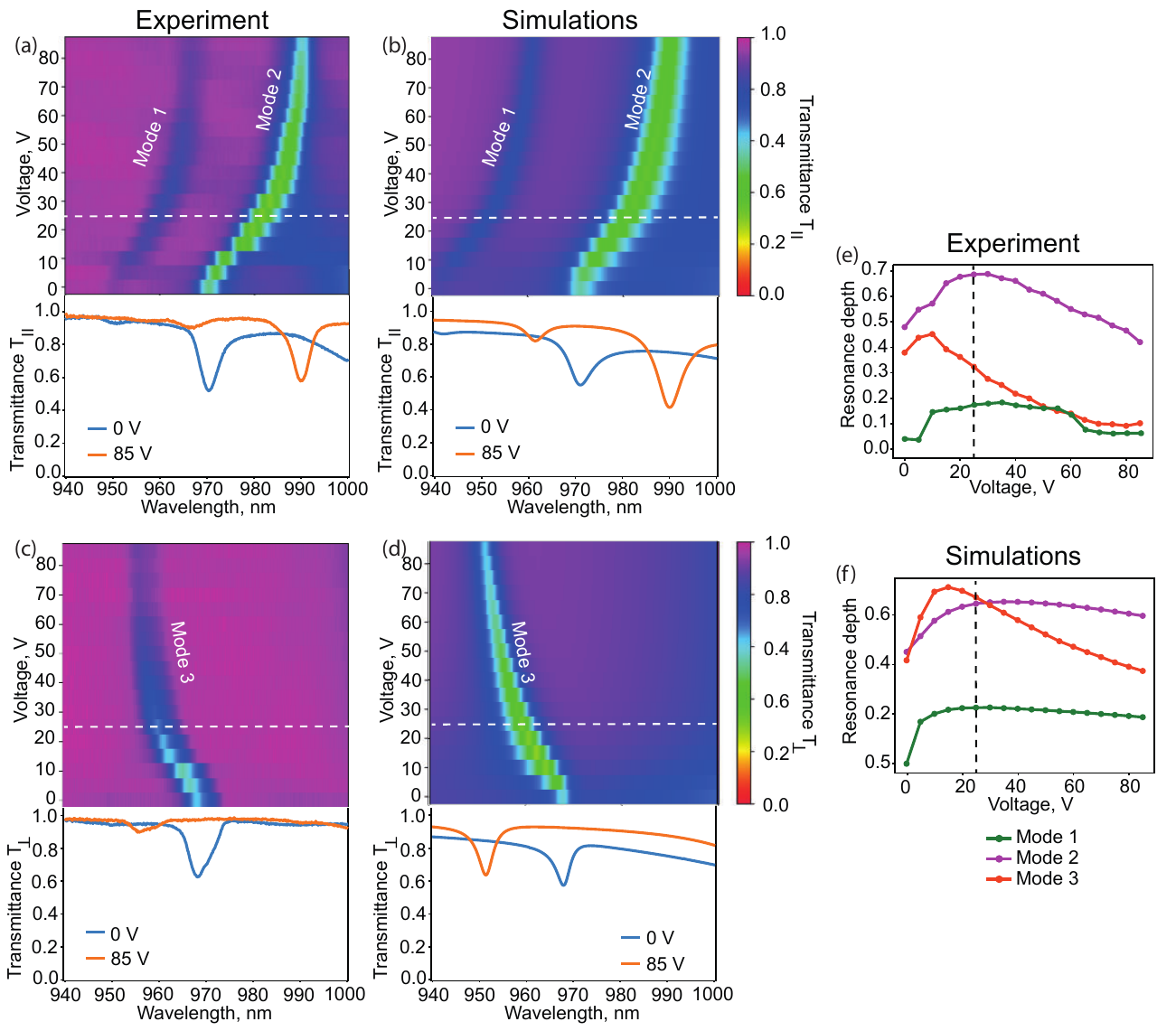}
  \caption{Measured and simulated transmittance spectra for (a,b) x-polarized ($T_{\parallel}$) and (c,d) y-polarized ($T_{\perp}$) incident light and for voltages ranging from 0~V to 85~V. (e) Measured and (f) simulated resonance depths as a function of the applied voltage.}
  \label{fig4}
\end{figure}

Next, making use of the multiresponsivity of the LC, we experimentally investigated the dynamic tunability of far-field coupling strength of the qBICs modes hosted by the metasurfaces.
First, to identify the conditions under which the asymmetry factor and, consequently, the far-field coupling of the hybrid metasurface are maximized, we applied a voltage (AC 1kHz) across the LC cell, which was varied in the range from 0 V to 85 V (Fig. \ref{fig4}). To this end, the LC cell was connected to a function generator (HM8030-4 from Hameg) and to a voltage amplifier (WMA-300 from Falco Systems). 
The measured voltage-dependent transmittance spectra for the two orthogonal linear polarizations are shown in Fig. \ref{fig4} (a,c).
Increasing the applied voltage from 0~V to 85~V results in a redshift of approximately 20~nm for Modes 1 and 2, while Mode 3 undergoes a blueshift of approximately 20~nm.\\
Furthermore, we performed full-wave numerical simulations of the hybrid system using the method by Zou et al.\cite{Zou2019}. The out-of-plane rotation of the LCs was calculated using equation (2) in the Supporting Information, which was derived by Alex-Amor et al. \cite{alex2019analytical}, with material parameters of the LC E7 \cite{chen2015simple, bankova2024optical}. For a coverplate plate without metasurface and coated with a photoalignment layer, an infinite anchoring energy and consequently an out-of-plane rotation of $\Theta_p=0 ^\circ $ at the surface \cite{liang2014analytical} were assumed (for further explanation see Section S2 in Supporting Information).
In our model, the pretilt of the LC molecules in the region next to the meta-atom is varied if voltage is applied to the LC cell. Thus, the LC molecules exhibit an out-of-plane rotation $\overline{\Theta_1}$ in this region. $\overline{\Theta_1}$ is the average of $\Theta_1$ over the height of the meta-atoms (see equation (6) in Supporting Information Section S2). $\Theta_1$ corresponds to equation (2) in the Supporting Information \cite{alex2019analytical} and is consequently a function of $\Theta_p1$ and the applied voltage $V$. 
These results are displayed in Fig. \ref{fig4}(b,d), showing a good qualitative agreement with the experimentally measured spectra (for detailed description of models see Supporting Information Section S2).
In the experimental spectra, we observe a significant reduction of the Q-factors of all three modes, which can be explained by fabrication imperfections \cite{Kuehne2021} and local variations in the LC orientation \cite{Dolan2021} resulting in diffuse scattering. 
Importantly, as the overall Q-factor is dominated by these imperfections and considering that the inverse of the total Q-factor is formed as the sum of the inverses of the Q-factors associated with the individual loss channels \cite{barreda2022impact, koshelev2019nonlinear}, we cannot expect a notable effect of the asymmetry-induced far-field coupling on the overall Q-factor in this regime.
 As a result, in the presence of dominating parasitic loss channels, which is the common regime for experimentally fabricated very-high-Q metasurface designs, the Q-factor is not a useful metric for quantifying the coupling strength between the qBICs and the radiation continuum.
To identify the voltage value, under which the radiation continuum is maximized in the experiment despite this limitation, we resort to use the resonance depth as an observable indicative of the coupling strength. The resonance depth directly correlates with the asymmetry in the presence of dominating parasitic loss channels (see Section S4 in Supporting Information for details). 
These losses can be modelled by adding a fictitious imaginary part $\kappa$ to the refractive index of the silicon meta-atoms \cite{benisty2000radiation}. A loss of $\kappa=0.003$ was added for the metasurface in air and a loss of $\kappa=0.018$ for the metasurface embedded in a LC cell.
The measured and simulated resonance depths as a function of applied voltage extracted from the data displayed in Fig. \ref{fig4} (a-d) are shown in Fig. \ref{fig4} (e,f). For all modes, the experimental resonance depth initially increases with increasing voltage and then decreases again. Overall, an applied voltage of 25~V yields the deepest resonance depths averaged over all three modes (dashed vertical line in Fig. \ref{fig4} (e,f)) and is chosen for our central experiments in the following. The simulated resonance depths show qualitatively similar behaviour. Differences in the exact resonance depths and their evolution with voltage are likely due to the simplifications made in the model regarding the complexities of LC alignment in the vicinity of the meta-atoms. 

\begin{figure}
\centering
\includegraphics[trim=1 1 1 1,clip,width=\textwidth]{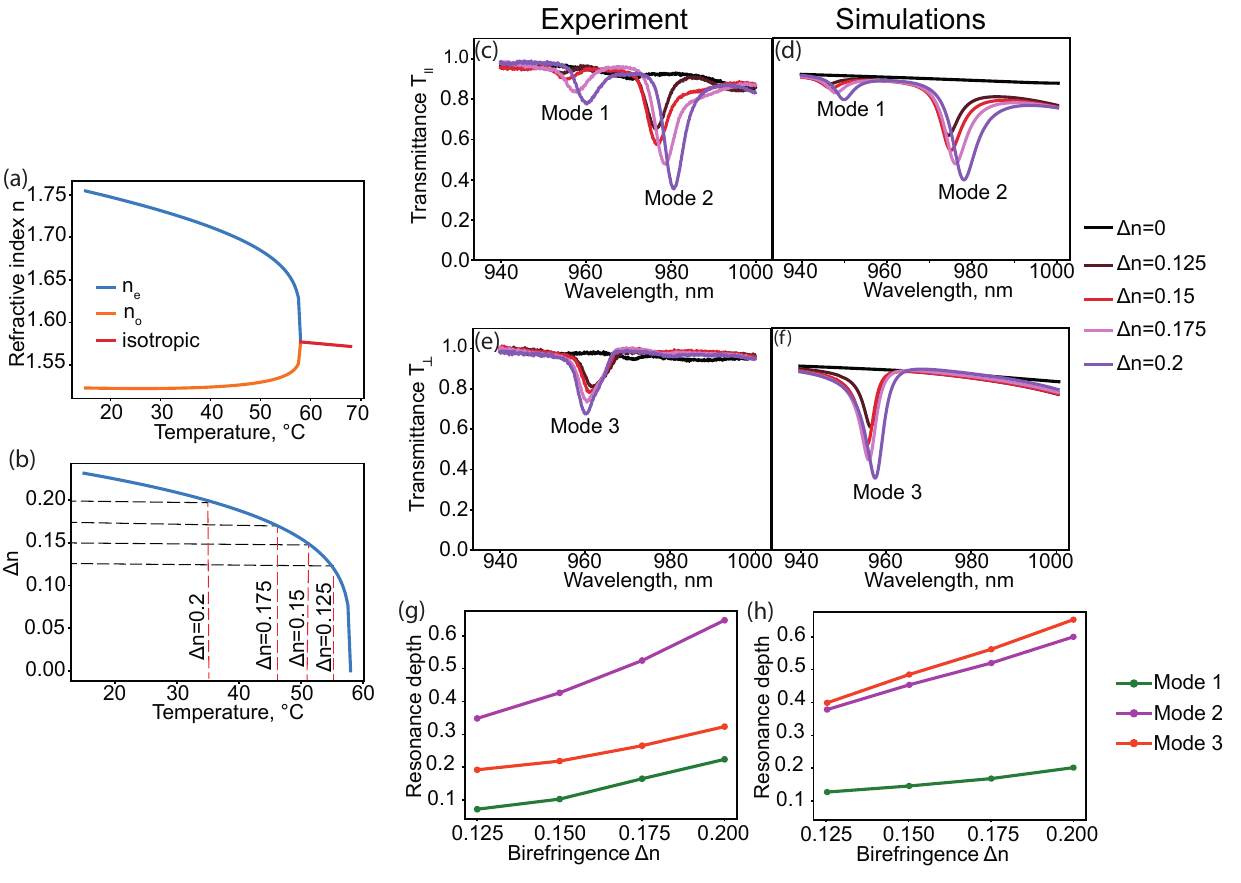}
  \caption{(a) Temperature-dependent refractive indices and (b) birefringence $\Delta$n for liquid crystal E7 at a wavelength of 950~nm. These spectra represent interpolated data from Li et al. \cite{Li2005}. Measured and simulated transmittance spectra for (c, d) x-polarized ($T_{\parallel}$) and (e, f) y-polarized ($T_{\perp}$) incident light, for $\Delta n=0$ and for birefringence values ranging from $\Delta n=0.125$ to $n=0.2.$ in steps of 0.025 at an applied voltage of 25~V. (g) Measured and (h) simulated resonance depths extracted from (c-f).}
  \label{fig5}
\end{figure}

Next, to investigate the occurrence and properties of the qBICs as a function of anisotropy induced asymmetry, we heated the LC cell while applying the optimal voltage value of 25~V.  
Fig. \ref{fig5} (a) shows the ordinary ($n_o$) and extraordinary refractive index ($n_e$), and Fig. \ref{fig5} (b) the birefringence $\Delta n=n_e-n_o$ of the liquid crystal E7 as a function of temperature at a wavelength of 950~nm.
The displayed data is based on the publication by Li et al. \cite{Li2005}. The interpolation of the data was performed using the package NumPy in Python.
Importantly, for temperatures exceeding the critical temperature of 58~$^\circ C$, the LC transitions into its isotropic state, which was previously shown to restore an isotropic dielectric environment for integrated metasurfaces \cite{Sautter2015, Zou2021, Beddoe2025}.  
To investigate the dependence of the qBIC behaviour on the anisotropy, it is necessary to measure and simulate the qBIC response for varying values of the birefringence $\Delta n$. Thus, in addition to the isotropic state, we consider birefringence values, $\Delta n$, ranging from 0.125 to 0.20 in increments of 0.025.
Consequently, temperatures of 55~$^\circ$C, 50~$^\circ$C, 44~$^\circ$C and 32~$^\circ$C were chosen. 
The temperature-induced change in the refractive index of silicon (approximately 0.01)  \cite{Rahmani2017} and the electro-optic Kerr effect in the isotropic phase of E7, which causes a birefringence of $\Delta n \approx 0.001$ \cite{gilani2001electro}, can both be neglected.
Fig. \ref{fig5} (c-f) compares the simulated and measured transmittance spectra for incident x-polarized light ($T_{\parallel}$, Fig. \ref{fig5} (c,d)) and y-polarized ($T_{\perp}$, Fig. \ref{fig5} (e, f)). As an important observation, at $\Delta n=0$, corresponding to the isotropic state of the LC, the sharp resonant features associated with the qBICs disappear. This clearly indicates that the qBICs are induced by the anisotropy of the LC.

For finite anisotropy values, the resonance depths increase monotonously with the birefringence for the measured (Fig. \ref{fig5} (g)) and simulated (Fig. \ref{fig5} (h)) spectra. These results demonstrate that nematic LCs provide a viable platform for dynamic control of the coupling of the qBICs to the radiation continuum using voltage and temperature as control parameters. Importantly, this coupling cannot only be tuned but also be completely switched off owing to the transition of the LC in its isotropic state.

\section{Conclusions and Outlook}
We demonstrated that qBICs can be excited in C$_{4v}$- symmetric metasurfaces composed of silicon nanocylinders arranged on a square lattice by embedding them into an anisotropic birefringent optical medium. Moreover, we showed that their coupling strength to the radiation continuum can be tuned by changing the magnitude of the birefringence and the orientation of the anisotropy axis of the medium. Experimentally, we realized such a system by embedding a fabricated silicon nanocylinder metasurface into a nematic LC cell and leveraging the multiresponsivity of the LCs. Specifically, we controlled the anisotropy axis of the LC by applying voltage to the LC cell and changed the magnitude of the anisotropy by controlling the temperature of the LC. Observing the resonance depths of the qBICs as a function of applied voltage, we identified a voltage of 25~V to maximize anisotropy-induced coupling to the radiation continuum. When fixing the applied voltage at this value, the resonance depth, which is associated to the coupling strength to the continuum, could be systematically tuned by the magnitude of the birefringence of the LC via temperature. Importantly, for temperatures exceeding the critical value, when the LC transitions to its isotropic phase, the qBIC signatures vanish completely. Our findings do not only shed new light on the concept of anisotropy-induced qBICs, but also open new avenues for actively reconfigurable metasurfaces and metadevices. In particular, the capability to switch resonances completely off is in stark contrast to common resonance tuning effects, thereby providing interesting new engineering degrees of freedom. Moreover, the demonstrated active tunability of far-field coupling strength to the continuum is of particular interest in highly integrated active metasystems, such as for tunable light-emitting, lasing and nonlinear metasurfaces, whose modal dynamics can sensitively depend on far-field coupling strength. Such tunable active metasystems offer interesting opportunities for applications in displays, optical modulation and tunable photonic devices.

\section{Author Information}
\subsection{Funding Sources}
This work was funded by the Deutsche Forschungsgemeinschaft (DFG, German Research Foundation), Project-ID: 437527638 -- IRTG 2675 (Meta-Active). This project was made possible by funding from the Carl Zeiss Foundation. It was supported by the Free State of Thuringia and the European Social Fund Plus. A.B. gratefully acknowledges financial support from Spanish national project No. PID2025-169616NB-C44. A.B. thanks MICINN for the Ramon y Cajal Fellowship (grant No. RYC2021-030880-I). 

\begin{suppinfo}
Transmission spectra of the sample in air; Numerical models; Near-field maps of Modes 1 to 3; Resonance depth as a function of the asymmetry parameter $\alpha$; Experimental setup for transmittance measurements; Construction process of the LC cell 
\end{suppinfo}

\bibliography{literaturphd}

\end{document}


\section{S1. Transmittance spectra of the sample in air}

\begin{figure}[H]
\includegraphics[trim=1 1 1 1,clip,width=\textwidth]{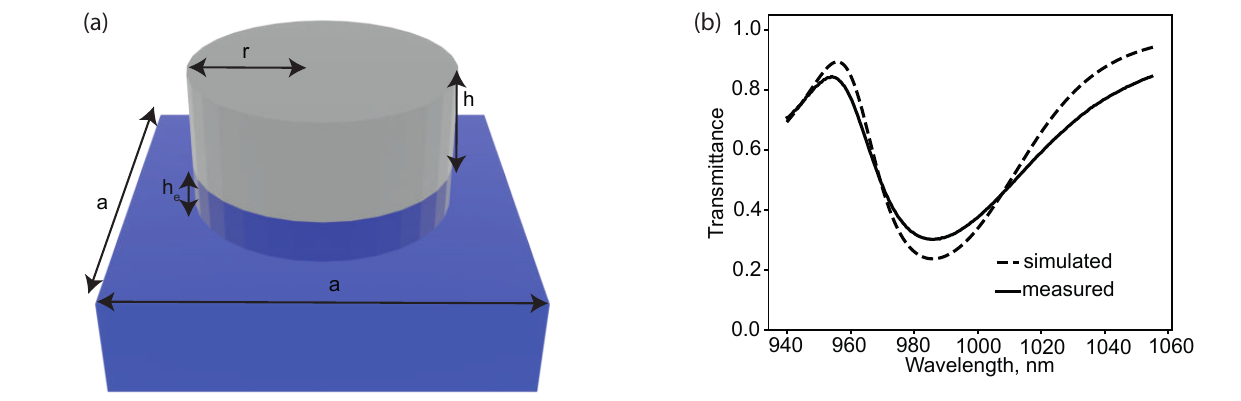}
  \caption{(a) Sketch of the metasurface unit cell with period of $a~=~520~nm$, radius $r~=~179~nm$, height $h~=~166~nm$ and overetch $h_e~=~53~nm$. (b) Measured and simulated transmittance spectra of the metasurface in air.}
  \label{fS2}
\end{figure}

Figure \ref{fS2} shows a sketch of the unit cell of the metasurface and a comparison of the simulated and measured transmittance spectrum for the metasurface. The resonance is a Huygens' resonance \cite{Decker2015}.

\section{S2. Numerical models}
\begin{figure}[htbp]
\centering
\includegraphics[trim=1 1 1 1, clip, width=\textwidth]{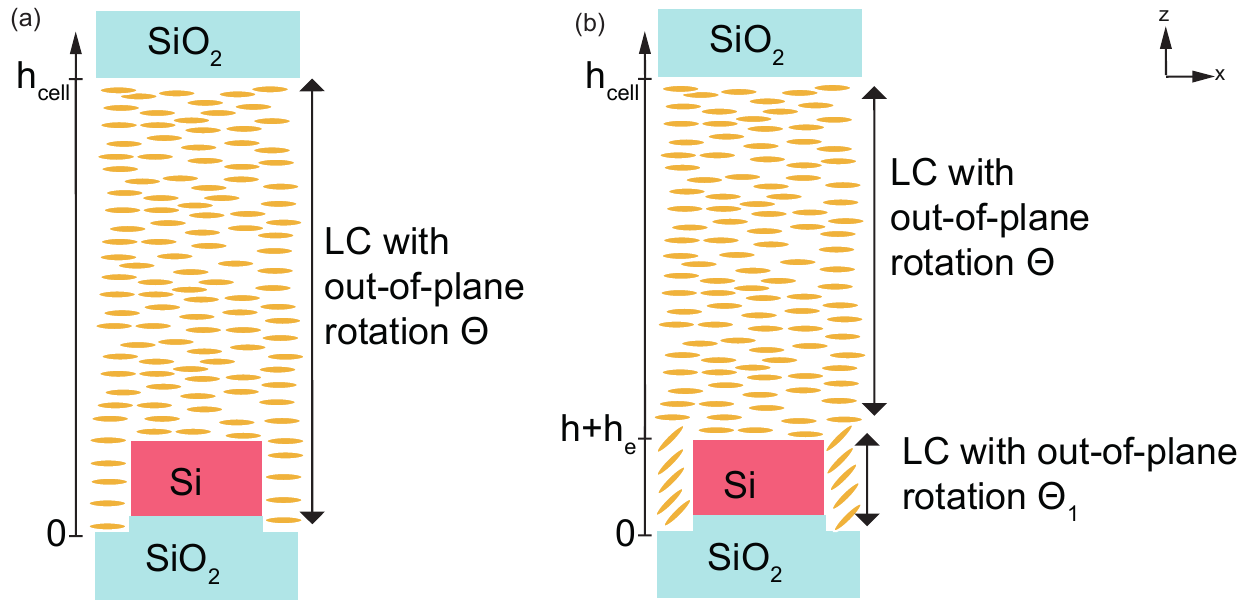}
\caption{ Two different models for simulating metasurfaces embedded in LC cells: (a) Ideal model and (b) refined model. $h+h_e$ is the total height of the meta atom. $h_{cell}$ is the thickness of the LC cell}
\label{modellcmeta}
\end{figure}

This section describes the numerical models and settings used in the simulations. The calculations were performed using the Frequency Domain Solver in CST Studio Suite and the finite-difference time-domain (FDTD) solver in Lumerical. Simulating periodic metasurfaces requires specific settings in both software packages, including periodic boundary conditions along the x-z and y-z planes.
The boundaries are set to glass coverplates with a refractive index of 1.45 \cite{malitson1965interspecimen}.
Furthermore, a planewave source is used as a source for the incident light.
The LCs are implemented as an anisotropic material with its slow axis oriented x-direction and tilted by the out-of-plane rotation $\Theta$.
Figure \ref{modellcmeta} shows the two different ways to model a LC-based metasurface.
In both models, the unit cell consists of two $SiO_2$ layers for the coverplate and the substrate of the metasurface and the silicon metasurface. The main difference between the models is the orientation of the LC molecules.
In the model shown in Fig. \ref{modellcmeta} (a), the ideal model, the orientation of the LC molecule is homogeneously distributed over the LC cell. This model was used in Figures 2 and 3 of the main manuscript. However, this does not represent a realistic situation. Due to an equilibrium of the anchoring energies of the photoalignment material \cite{muravsky2024q} and the meta-atoms \cite{muravsky2024q, Veetil2024}, the LC molecules pretilt in the region next to the metasurface \cite{Zou2019}. However, no exact physical model is developed to simulate the exact out-of-plane rotation of the LC molecules in the region close to the metasurface. A simplified way of modelling this effect is the refined model shown in Figure \ref{modellcmeta} (b) \cite{Beddoe2025}. In the region between the meta-atoms, the LC molecules exhibit a pretilt with an out-of-plane rotation of $\Theta_1$, the pretilt angle.
This model was used in Figures 1,4 and 5 in the main manuscript.

The transmission spectra for different voltages $V$ were simulated with the method by Zou et al. \cite{Zou2019}. The dependence of the out-of-plane rotation $\Theta(V, z)$ on the voltage $V$ and the position $z$ on z-axis was modelled with the equation from the work by Alex-Amor et al. \cite{alex2019analytical}, whereas $h_{cell}$ is the thickness of the LC cell.

\begin{equation} \Theta(V, z)= 2tan^{-1}( e^{z\sqrt{A(V)}+\Theta_p'})+2tan^{-1}( e^{-(z-d)\sqrt{A(V)}+\Theta_p'})  +c\label{eqLE2} \end{equation}
with 
\begin{equation} c=\Theta_p- 2tan^{-1}( e^{\Theta_p'})+2tan^{-1}( e^{d\sqrt{A(V)}+\Theta_p'})  \label{eqLE3} \end{equation}
,
\begin{equation}
\Theta_p'= ln(tan(\Theta_p)+sec(\Theta_p)) 
\label{eqLE4} \end{equation}
and 
\begin{equation}
A(V)=\frac{\epsilon_0 \Delta \epsilon (\frac{V}{h_{cell}})^2}{K_{33}} 
\label{eqLE5} \end{equation}

$\Theta_p$ is the out-of-plane rotation at the boundaries $z=0$ (Model in Figure \ref{modellcmeta}) and $z=h+h_e$ (Model in Figure \ref{modellcmeta} (b)) and $z=h_{cell}$ (both models). $h+h_e$ is the total height of the meta-atom with $h$ as the height of the silicon nanoresonator and $h_e$ the height of the overetch. The out-of-plane rotation depends on the anchoring energy of the photoalignment layer \cite{liang2014analytical}. 
Following values were used in our simulations: $\Delta \epsilon=14$, $K_{33}= 16 \cdot 10^{-12} $ and $\Theta_{p} =0 ^\circ$.
These are taken from Chen et al. \cite{chen2015simple} and Bankova et al.\cite{bankova2024optical}.
The pretilt angle in the vicinity of the metasurface was determined by fitting the real model to the experimental response at 0 V, yielding the out-of-plane rotation $\Theta_{P1}=38 ^\circ$ at the boundary $z=0$ in the real model (Figure \ref{modellcmeta} (b)). At 0 V, in the absence of electric-field-induced reorientation, the out-of-plane rotation in the pretilt region is therefore given by $\Theta_1 = \Theta_{P1}$. Upon application of a voltage V, the LC orientation in this region is also modified. This effect was taken into account by calculating the spatially dependent tilt angle $\Theta_1 (z)$ and averaging it over the vertical extent of the meta-atoms, from $z=0$ to $z=h+h_e$:
\begin{equation}
\overline{\Theta_1(z)} = \int_{0}^{h+h_e} dz\,\Theta_1(z)\left[\frac{1}{h+h_e}\right]
\label{eq:event}
\end{equation}

\section{S3. Near-field maps of Modes 1 to 3}
\begin{figure}[H]
\includegraphics[trim=1 1 1 1,clip,width=\textwidth]{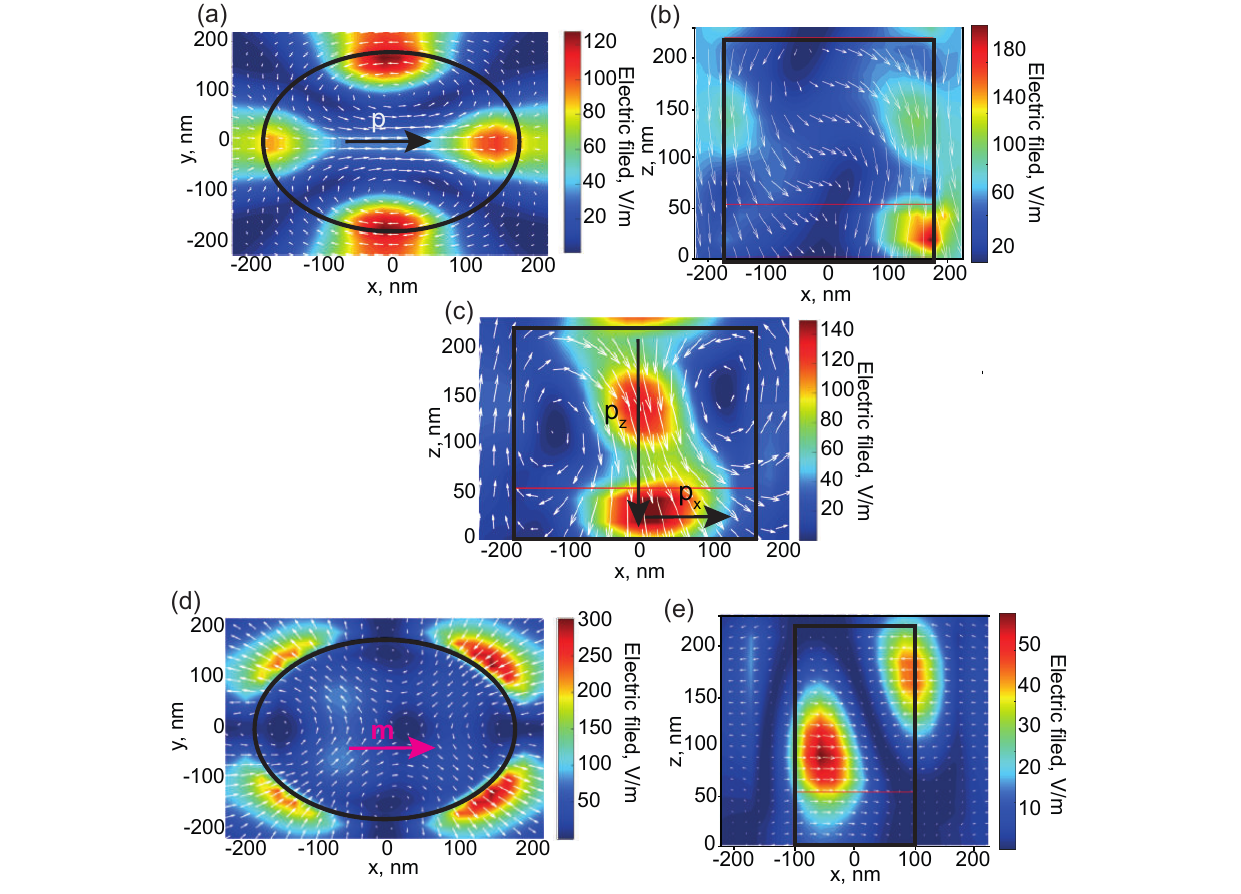}
  \caption{Mode 1: Near-field maps of the electric fields at the (a) x-y-plane (z~=~100~nm) and (b) x-z-plane (y~=~0~nm). 
  Mode 2: Near-field maps of the electric field at the (c) x-z-plane (y~=~0~nm). 
  Mode 3: Near-field maps of the electric field at the (d) x-y-plane (z~=~100~nm) and the (e) y-z-plane (x~=~75~nm). $p$ is the electric dipole moment. $p_x$ and $p_z$ are the electric dipole moments in x- and z-direction. $m$ is the magnetic dipole moment. The white arrows illustrate the polarization vectors of the fields.}
  \label{fS1}
\end{figure}

\section{S4. Resonance depth as a function of the asymmetry parameter $\alpha$}

\begin{figure}
\includegraphics[trim=1 1 1 1,clip,width=\textwidth]{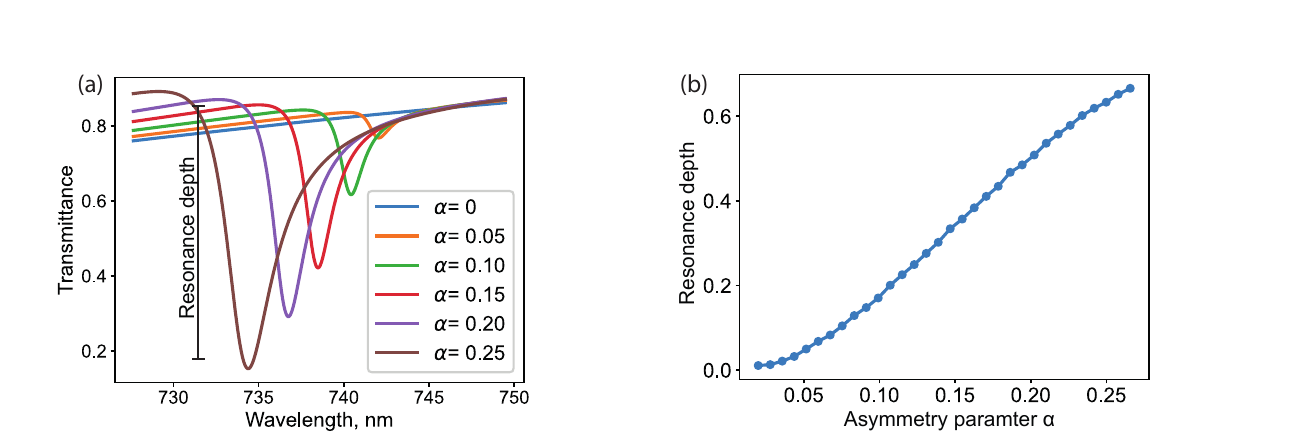}
  \caption{Simulated transmittance spectra for a metasurface with bars with a width of 112~nm, lengths from 221~nm to 301~nm and a height of 165~nm. (b) Simulated resonance depths for different asymmetry parameters $\alpha$}
  \label{figS3}
\end{figure}

As discussed in the main text, in the presence of optical losses, the total Q-factor is determined by the radiative and non-radiative decay channels and therefore does not provide a direct measure of the coupling strength between the qBIC and the radiation continuum \cite{Kuehne2021, barreda2022impact}. To further investigate the radiative coupling, we consider a metasurface with geometrical symmetry breaking, where the coupling of the qBIC to the radiation continuum is controlled by the asymmetry parameter $\alpha$ \cite{koshelev2018}.

Figure \ref{figS3} shows the transmission spectra for a qBIC-based metasurfaces for different asymmetry-parameters $\alpha$. The metasurface is based on silicon bars with a width of 112~nm, lengths from $l_1~=~221 nm$ to $l_2~=~301~nm$ and a height of 165~nm and a lattice constant of 440~nm \cite{Kuehne2021}. 
The asymmetry parameter $\alpha$ is calculated by the formula from Koshelev et al. \cite{koshelev2018}.
\begin{equation}
\alpha = \frac{|l_{1}-l_{2}|}{l_{2}},
\label{eq:asymmetry}
\end{equation}
To demonstrate the effect of $\alpha$ on the optical response of the metasurface with non-radiative decay channels, a loss of $\kappa~=~9.7 \cdot 10^{-3}$ is added to the refractive index of the silicon \cite{barreda2022impact}. 
In Figure \ref{figS3} (a), it can be seen that the resonances become more shallow for decreasing $\alpha$. This can be defined as the resonance depth (sketch in Figure \ref{figS3} (a)). The resonance depth directly correlates with $\alpha$ (Figure \ref{figS3} (b)). Thus, the resonance depth can be used to determine $\alpha$ in metasurfaces with a loss.

\section{S5. Experimental setup for transmittance measurements}
\begin{figure}
\includegraphics[trim=1 1 1 1,clip,width=\textwidth]{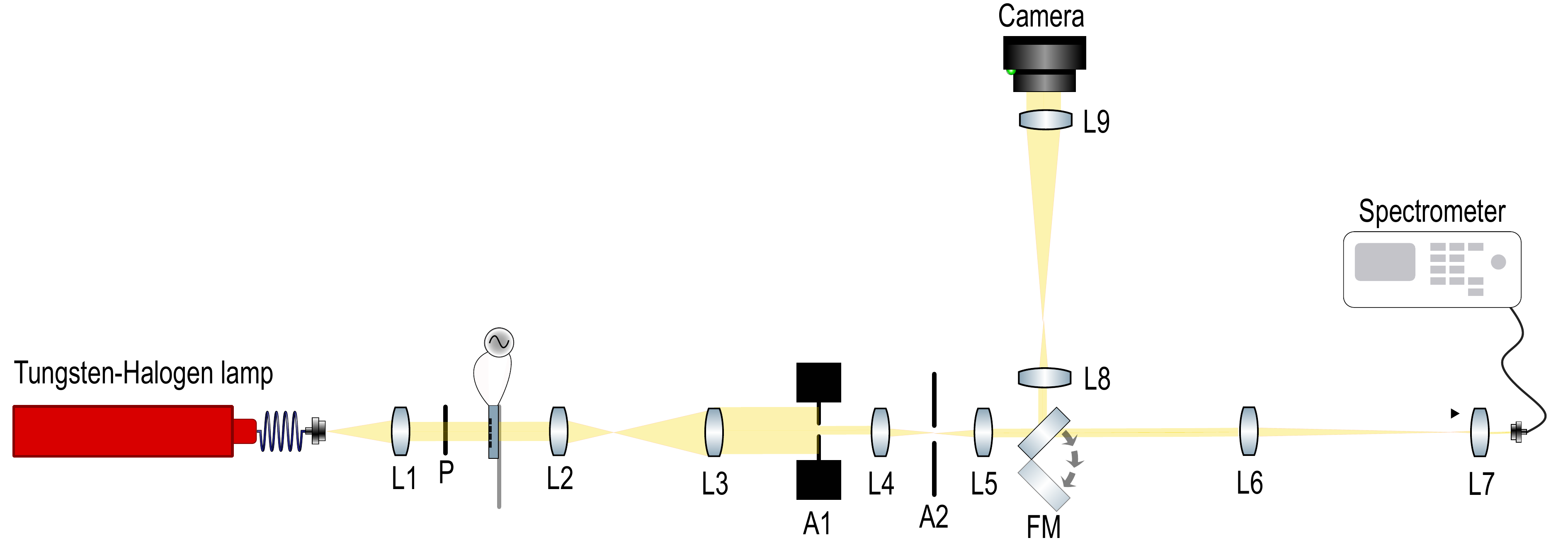}
  \caption{Sketch of the optical setup used for the transmittance measurements}
  \label{figS6}
\end{figure}
The transmittance measurements of the LC metasurface cell was performed with the custom built setup sketched in Fig. \ref{figS6}.
In this setup, a stabilized tungsten-halogen light source (Thorlabs SLS201L) was used as a light source. The outgoing beam was collimated with the lens L1 with a focal length of f$_{L1}$\, =\,7 \,mm. The polarization of the light is set with the linear polarizer (P), which can be mechanically rotated to vary the polarization of the light. The beam is then transferred to the sample and expanded by lenses L2 (f$_{L1}$\, = \, 50 \,mm) and L3 (f$_{L2}$\, = \, 200 \,mm). Aperture A1 is used to select the desired region on the sample for the measurement. Next, the image of the sample is transformed to k-space space by lense L4 (f$_{L4}$\,=\,50\,mm). Aperture A2 acts as a lowpass filter of the image. Lense L5 (f$_{L5}$\,=\,50\,mm) collimates the beam.
To find the right position on the sample, the sample should be imaged on a camera. Thereforefore, the beam is transferred by a flip mirror FM to a camera and expanded by lenses L8 (f$_{L8}$\,=\,50\,mm) and L9 (f$_{L9}$\,=\,200\,mm).
Finally, the diameter of the beam is reduced by lenses L6 (f$_{L3}$\,=\,100\,mm) and L7 (f$_{L7}$\,=\,50\,mm)  and coupled into the fibre of the spectrometer Yokogawa.

\section{S6. Construction process of the LC cell}

The LC cells in our sample consist of a SiO$_2$ glass cover plate and a metasurface sample with dimensions of 20~mm~×~30~mm. As a first step, the cover plate and the sample were cleaned with acetone, isopropanol, dimethyl formadide and de-ionized water. Afterwards, they were placed into an oxygen plasma cleaner for 5~minutes. This step is required to remove any organic contamination. 
Next, a solution of the photoalignment material was prepared by dissolving AmA2522 (in powder form) in a mixed solvent of water and isopropanole (volume ratio 1:1). The solution was then filtered using a 0.2~$\mu$m filter and spin-coated on the surface of the glass plate and on the metasurface sample. To this end, both were heated up using a hot plate for 2~minutes to $80\,^\circ$C and then quickly placed on the spin-coater. Spin coating was performed at a speed of 800~rpm for 5~s, followed by 3000~rpm for 40~s and finished with 800~rpm for 5~s. Next, the spin-coated sample and cover plate were baked on a hot plate for 5~minutes at $140\,^\circ$C. This process resulted in a photoalignment layer of approximately 15~nm thickness \cite{Zou2019}. 
To form the actual cell, the glass plate and the metasurface sample were pressed together using a metallic sample clip, which has a 20~mm~×~20~mm window.
The thickness of LC cell was determined by a transmittance spectrum of the empty cell fitted with a Fabry-Perot model \cite{Herna1986}. The determined thickness was used to estimate the volume of LC, which should be injected into the LC cell. The LC was injected with a pipette at the edge of the assembled cell. The capillary effect spreads the LCs evenly throughout.
Finally, the assembled and infiltrated cell was placed in the constructed DMD setup (see Ref. \cite{Beddoe2025}). The sample was exposed with y-polarized light with a wavelength of 457~nm at a power of 41~mW for 15~minutes. A cw laser with 457~nm was used as a source (Photontec 457~nm DPSS). The sample was heated during the exposure to bring the LC in its isotropic state. The necessary temperature increase above the critical temperature of the LC was achieved using a 15~$\Omega$ heating resistor attached to the backside of the sample and a PT100~RTD temperature sensor feedback system controlled with a temperature controller (LDT-5980, Newport).
Finally, the ITO electrodes were coated with a copper electrode.

\newpage
\bibliography{literaturphd}